\documentclass[twocolumn,
 groupedaddress, superscriptaddress,
 amsmath,amssymb,
 aps,
prl,lengthcheck,reprint]{revtex4-2}

\usepackage[utf8]{inputenc}
\usepackage[T1]{fontenc}
\usepackage{graphicx}
\usepackage{dcolumn}
\usepackage{bm}
\usepackage{hyperref}
\usepackage{xcolor}

\newcommand{\rs}{{\hat{\boldsymbol{R}}_\mathrm{S}}}
\newcommand{\ps}{{\hat{\boldsymbol{P}}_\mathrm{S}}}
\newcommand{\rn}{{\hat{\boldsymbol{r}}_n}}
\newcommand{\pn}{{\hat{\boldsymbol{p}}_n}}
\newcommand{\rzero}{{\hat{\boldsymbol{R}}_0}}
\newcommand{\comm}[2]{\left[#1,\,#2\right]}
\newcommand{\acomm}[2]{\left\{#1,\,#2\right\}}
\newcommand{\avg}[1]{\left\langle #1 \right\rangle}

\begin{document}

\preprint{APS/123-QED}

\title{Galilean boost invariance does not survive the trace: \\symmetry breaking in open quantum systems}

\author{Leonardo F. Calder\'on}
\affiliation{guane Enterprises, R+D+I Unit, Medell\'in 050010, Colombia.}

\author{Esteban Marulanda}
\affiliation{Grupo de F{\'i}sica At{\'o}mica y Molecular, Instituto de F{\'i}sica, Facultad de Ciencias Exactas y Naturales, Universidad de Antioquia, Medell\'in, Colombia.}
\affiliation{Instituto de Física, Universidade de São Paulo, 05508-090, São Paulo, SP, Brazil}

\author{Santiago Morales}
\affiliation{guane Enterprises, R+D+I Unit, Medell\'in 050010, Colombia.}
\affiliation{Grupo de F{\'i}sica At{\'o}mica y Molecular, Instituto de F{\'i}sica, Facultad de Ciencias Exactas y Naturales, Universidad de Antioquia, Medell\'in, Colombia.}

\author{Leonardo A. Pach\'on}
\affiliation{guane Enterprises, R+D+I Unit, Medell\'in 050010, Colombia.}

\date{\today}

\begin{abstract}
Tracing out a Galilean-invariant Caldeira--Leggett environment breaks Galilean boost covariance of the reduced dynamics, while spatial translations and rotations survive intact. An operator-level analysis of the exact Hu--Paz--Zhang master equation localizes the violation entirely in the dissipative anticommutator term, scaling with the damping coefficient $\Gamma(t)f(t)$. The fluctuation--dissipation theorem ties this coefficient to the absorptive bath response that drives equilibrium momentum diffusion, so for any non-trivial bath spectral density bilinear-coupled Galilean invariance, the fluctuation--dissipation theorem, and reduced boost covariance cannot hold simultaneously. The stochastic decomposition of the influence functional extends the mechanism beyond the quadratic regime. The dimensionless ratio $\hbar\gamma/k_\mathrm{B} T$ delineates the crossover: cold atoms in dissipative optical lattices and ultracold molecules sit at its edge. Parametric driving offers a one-directional escape: the squeezing rate that protects nonequilibrium entanglement above the standard quantum limit also suppresses boost-breaking over a driving cycle.
\end{abstract}

\maketitle

\emph{Introduction}\textemdash Approximations governing the dynamics of a quantum system interacting with its environment often elude independent assessment. When the microscopic details of the environment remain unknown, symmetry principles provide one of the few systematic guidelines for constraining the effective dynamics. Several contexts---environment-induced superselection, virtual subsystems, quantum reference frames---establish that the partial trace can break the global symmetries of a composite system~\cite{Bartlett2007, Zanardi2001, 2007a}. A complementary line of inquiry asks when dynamical maps governing the reduced evolution remain \emph{covariant} under the Galilean group~\cite{PhysRevLett.119.100403, Holevo1996}.

Holevo characterized the most general Markovian, completely positive, trace-preserving maps that are translation and boost covariant~\cite{Holevo1993a, Holevo1993b, Holevo1995, Holevo1996}, and Gasbarri, Toro\v{s}, and Bassi extended the classification to the non-Markovian Gaussian case~\cite{PhysRevLett.119.100403}. Covariance constraints have since played a central role in the collapse-model and macroscopicity literature~\cite{Bassi2013}. These developments characterize the structure that effective dynamical maps must possess once Galilean covariance is imposed, motivated by the expectation that space-time symmetries holding at the fundamental level transfer to the reduced level.

This Letter investigates the complementary microscopic question: does Galilean covariance survive the partial trace over a manifestly Galilean-invariant environment, and---when it does not---which operator structure carries the breaking? Already for a free particle undergoing unitary evolution, the Liouville--von Neumann equation fails to transform covariantly under boosts in the standard sense---it remains \emph{invariant} only when the Hamiltonian transforms appropriately~\cite{Lombardi2010, Ballentine1998-nt}. The partial trace over environmental degrees of freedom introduces a qualitatively different mechanism for symmetry breaking: the generators of the Galilean group for the composite system involve environmental operators that disappear once the trace removes them.

The Caldeira--Leggett model provides the natural analytical platform for this inquiry: it is the most general exactly solvable framework for quadratic system--bath couplings~\cite{CaldeiraLeggett1983, Hakim1985}, underlies the broader class of quadratic-map applications, and admits an exact reduced master equation~\cite{Hu1992} that opens the operator structure of the dissipative dynamics to direct inspection. Within this setting, the exact Hu--Paz--Zhang (HPZ) master equation enables an operator-level identification of the \emph{dissipative anticommutator} as the unique boost-breaking term, while the Hamiltonian, anomalous-diffusion, and normal-diffusion terms transform covariantly. A second extension, central to the present work, follows from the stochastic decomposition of the influence functional~\cite{Stockburger2002, Koch2008, Diosi1998}: the boost-violation mechanism survives the passage to arbitrary system potentials and so generalizes beyond the quadratic regime. Although the HPZ master equation has been analyzed extensively for decoherence, classicality, and Lindblad reduction~\cite{Halliwell1996, AnastopoulosHu2000, FlemingHu2012, Anglin1997}, the action of the Galilean group on the reduced master equation has not previously been addressed. The operator-level analysis below isolates dissipation---rather than decoherence or diffusion---as the physical driver of Galilean violation and identifies the experimental regimes where covariant maps cease to admit a microscopic foundation.

\emph{Lagrangian formulation and Galilean invariance}\textemdash The Lagrangian formulation places the discussion on firm ground. Consider a system particle of mass $M_\mathrm{S}$ coupled to $N$ bath oscillators with masses $m_n$ and frequencies $\omega_n$. The most general quadratic Lagrangian whose interaction terms depend only on coordinate differences (a necessary condition for Galilean invariance) reads
\begin{equation}\label{eq:lag1}
L = \frac{1}{2}M_\mathrm{S}\dot{\boldsymbol{q}}_0^{2} + \frac{1}{2}\sum_{n=1}^{N} m_n \left(\dot{\boldsymbol{q}}_n^2 - \omega_n^2 (\boldsymbol{q}_0 - \boldsymbol{q}_n)^2\right) - V(\boldsymbol{q}_0),
\end{equation}
where $\boldsymbol{q}_0$ denotes the classical system coordinate, $\boldsymbol{q}_n$ ($n=1,\ldots,N$) the bath coordinates, and $V(\boldsymbol{q}_0)$ an external potential acting on the system. For a harmonic potential $V(\boldsymbol{q}_0) = \frac{1}{2}M_\mathrm{S}\omega_\mathrm{S}^2(\boldsymbol{q}_0 - \boldsymbol{R}_0)^2$, the Lagrangian reduces to the Caldeira--Leggett model. Under a Galilean boost $\boldsymbol{q}_j \to \boldsymbol{q}_j + \boldsymbol{v}\,t$ for all $j$ (including the system), the kinetic terms shift by total time derivatives, while the interaction terms $(\boldsymbol{q}_0 - \boldsymbol{q}_n)^2$ remain manifestly invariant. Explicitly,
\begin{equation}\label{eq:boost_lag}
L' = L + \frac{\mathrm{d}}{\mathrm{d}t}\!\left[\sum_{j=0}^{N} m_j\,\boldsymbol{q}_j\cdot\boldsymbol{v} + \tfrac{1}{2}M\,\boldsymbol{v}^{2}\,t\right],
\end{equation}
where $m_0 \equiv M_\mathrm{S}$ and $M = M_\mathrm{S} + \sum_n m_n$, so the action changes by a boundary term and the equations of motion remain unchanged. This establishes the Galilean invariance of the full model at the classical level. A strictly invariant variant of the Lagrangian and the standard regularization of the free-particle limit $\omega_\mathrm{S}\to 0$ admit the same structural conclusions and appear in the Supplemental Material (SM)~\cite{SM}.

The reference point $\boldsymbol{R}_0$ co-boosts throughout, $\boldsymbol{R}_0\to\boldsymbol{R}_0+\boldsymbol{v}\,t$, so that the relative coordinate $\boldsymbol{q}_0-\boldsymbol{R}_0$ remains Galilean-invariant. This is an idealization---a lab-frame optical trap, for instance, breaks Galilean symmetry trivially through its anchoring to the laboratory---adopted here to isolate the bath-induced mechanism of symmetry breaking from such trivial breaking.

\emph{Conserved currents}\textemdash By Noether's theorem, the boost invariance of the Lagrangian~\eqref{eq:lag1} implies the conservation of the boost generator
\begin{equation}\label{eq:noether}
\boldsymbol{G} = \sum_{j=0}^{N} m_j \boldsymbol{q}_j - \boldsymbol{P}_{\mathrm{tot}}\, t,
\end{equation}
where $\boldsymbol{P}_{\mathrm{tot}} = \sum_j m_j \dot{\boldsymbol{q}}_j$ is the total momentum. Its quantum counterpart plays a central role in the analysis that follows.

\emph{Invariance of the global quantum dynamics}\textemdash Quantization promotes the classical coordinates $\boldsymbol{q}_0$ and $\boldsymbol{q}_n$ to operators $\rs$ and $\rn$, with conjugate momenta $\ps$ and $\pn$. The Hamiltonian corresponding to the Lagrangian~\eqref{eq:lag1} with a harmonic system potential reads
\begin{equation}\label{eq:H}
\begin{split}
\hat{H} = &\frac{\ps^{2}}{2M_\mathrm{S}}+\frac{M_\mathrm{S} \omega_\mathrm{S}^{2}}{2}(\rs-\rzero)^{2}
\\ &+\sum_{n=1}^{N}\left[\frac{\pn^{2}}{2m_n}+\frac{m_n \omega_n^{2}}{2}(\rn-\rs)^{2}\right].
\end{split}
\end{equation}

Galilean invariance of the quantum theory requires the von Neumann equation to retain its form in every inertial frame~\cite{Lombardi2010}. For a unitary transformation $\hat{U}_g$ with $\hat{\rho}' = \hat{U}_g \hat{\rho}\,\hat{U}_g^\dagger$, the transformed equation reads
\begin{equation}
\label{eq:vN_transf}
\frac{\mathrm{d} \hat{\rho}'}{\mathrm{d} t}=-\mathrm{i}\comm{\hat{H}'}{\hat{\rho}'}+\frac{\mathrm{d}\hat{U}_g}{\mathrm{d}t}\hat{U}_g^{\dagger} \hat{\rho}'+\hat{\rho}'\hat{U}_g\frac{\mathrm{d}\hat{U}_g^{\dagger}}{\mathrm{d}t},
\end{equation}
with $\hat{H}' = \hat{U}_g \hat{H}\hat{U}_g^\dagger$. Invariance holds if and only if the last two terms cancel the difference $\hat{H}' - \hat{H}$ inside the commutator. For time-independent transformations this requires $\hat{H}' = \hat{H}$; for the time-dependent boost it requires $\hat{H}' = \hat{H} - \boldsymbol{u}\cdot\hat{\boldsymbol{P}}_\mathrm{T} + \text{const}$. The SM~\cite{SM} carries out the verification subgroup by subgroup. The result: the Hamiltonian~\eqref{eq:H} satisfies invariance for every element of the Galilean group, provided the generators are those of the \emph{total} (system + environment) degrees of freedom.

\emph{Master equation for the reduced dynamics}\textemdash Tracing over the bath yields the dynamics of $\hat{\rho}_\mathrm{S} = \mathrm{Tr}_\mathrm{B}[\hat{\rho}]$. With a factorized initial condition $\hat{\rho}(0) = \hat{\rho}_\mathrm{S}(0)\otimes\hat{\rho}_\mathrm{B}(0)$, the exact reduced master equation derived by Hu, Paz, and Zhang~\cite{Hu1992}---later recast in operator form by Schlosshauer~\cite{2007a}---reads
\begin{equation}\label{eq:master}
\begin{split}
\frac{\mathrm{d}\hat{\rho}_\mathrm{S}}{\mathrm{d}t} &= \hat{\mathfrak{L}}_t\,\hat{\rho}_\mathrm{S}(t) \\
&= -\mathrm{i}\comm{\frac{\ps^{2}}{2M_\mathrm{S}}+\kappa(t)(\rs-\rzero)^{2}}{\hat{\rho}_\mathrm{S}} \\
&\quad +\Gamma(t)\comm{\rs}{\comm{\ps}{\hat{\rho}_\mathrm{S}}} \\
&\quad -M_\mathrm{S}\Gamma(t)h(t)\comm{\rs}{\comm{\rs}{\hat{\rho}_\mathrm{S}}} \\
&\quad -\mathrm{i}\Gamma(t)f(t)\comm{\rs}{\acomm{\ps}{\hat{\rho}_\mathrm{S}}},
\end{split}
\end{equation}
where $\kappa(t) = \frac{1}{2}M_\mathrm{S}(\omega_\mathrm{S}^2 + \delta\Omega^2(t)) + \tilde{\Omega}^2$, with the frequency renormalization $\tilde{\Omega}^2 = \int d\omega\, I(\omega)/\omega$ fixed by the spectral density $I(\omega) = \sum_n m_n\omega_n^3\,\delta(\omega - \omega_n)$. The time-dependent coefficients $\delta\Omega^2(t)$, $\Gamma(t)$, $h(t)$, and $f(t)$ encode all environmental effects~\cite{Hu1992}.

The four lines of Eq.~\eqref{eq:master} play distinct physical roles. The first captures the unitary part with a renormalized frequency. The second ($\propto\Gamma$, mixed double commutator $[\rs,[\ps,\cdot]]$) represents the \emph{anomalous diffusion} term and describes cross-correlations between position and momentum. The third ($\propto\Gamma h$, double commutator with $\rs$) represents the \emph{normal diffusion} term: in the Wigner representation it generates diffusion in momentum, and in the position basis it serves as the primary source of decoherence, since $\langle x|\comm{\rs}{\comm{\rs}{\hat\rho_\mathrm{S}}}|x'\rangle=(x-x')^{2}\rho_\mathrm{S}(x,x')$. The fourth ($\propto\Gamma f$, commutator-anticommutator with explicit $\mathrm{i}$ ensuring hermiticity preservation) represents the \emph{dissipative} term responsible for damping: in the Markovian late-time limit, $\avg{\ps(t)} \sim e^{-\Gamma f\,t}\avg{\ps(0)}$.

\emph{Invariance of the reduced dynamics}\textemdash The relevant generators after the trace belong to the \emph{isolated} system: the dressed Hamiltonian $\hat{H}_\mathrm{S}$ defined by the first line of Eq.~\eqref{eq:master}, $\ps$, $\hat{\boldsymbol{J}}_\mathrm{S} = \rs \times \ps$, and $\hat{\boldsymbol{G}}_\mathrm{S} = M_\mathrm{S}\rs - \ps t$. The classical Noether charge $\boldsymbol{G}$ in Eq.~\eqref{eq:noether} sums over all degrees of freedom and is therefore extensive in the total mass $M=M_\mathrm{S}+\sum_n m_n$; its kinematic projection $\hat{\boldsymbol{G}}_\mathrm{S}$ onto the system Hilbert space carries only the fraction $M_\mathrm{S}/M$ of the total boost. No operator can act on $\hat{\rho}_\mathrm{S}$ with the full $\hat{\boldsymbol{G}}_\mathrm{T}$, because $\hat{\boldsymbol{G}}_\mathrm{T}$ contains bath degrees of freedom that the trace has already eliminated.

The transformation of the master equation under $\hat{U}_g$ takes the form
\begin{equation}\label{eq:master_transf}
\frac{\mathrm{d}\hat{\rho}_\mathrm{S}'}{\mathrm{d}t} = \hat{\mathfrak{L}}_t'\,\hat{\rho}_\mathrm{S}'(t) + \frac{\mathrm{d}\hat{U}_g}{\mathrm{d}t}\hat{U}_g^{\dagger}\hat{\rho}_\mathrm{S}' + \hat{\rho}_\mathrm{S}'\hat{U}_g\frac{\mathrm{d}\hat{U}_g^{\dagger}}{\mathrm{d}t},
\end{equation}
where $\hat{\mathfrak{L}}_t'$ denotes the superoperator~\eqref{eq:master} after substituting transformed counterparts for all operators. Invariance requires $\frac{\mathrm{d}\hat{\rho}_\mathrm{S}'}{\mathrm{d}t} = \hat{\mathfrak{L}}_t\,\hat{\rho}_\mathrm{S}'(t)$.

\emph{Spatial translations and rotations}\textemdash Under $\hat{U}_a = e^{-\mathrm{i}\boldsymbol{a}\cdot\ps}$, $\rs \to \rs - \boldsymbol{a}$ and $\ps \to \ps$. Since both $\rs$ and $\rzero$ shift by the same vector and commutators remain invariant, the master equation is \textbf{translation-invariant}. Under an infinitesimal rotation, expanding the terms in Eq.~\eqref{eq:master} to order $\theta$ shows that each pair of cross terms cancels by the Jacobi identity and the antisymmetry of the cross product. The master equation is therefore \textbf{rotation-invariant}.

\emph{Time translations}\textemdash Under $\hat{U}_\tau = e^{-\mathrm{i}\tau\hat{H}_\mathrm{S}}$ generated by the dressed Hamiltonian $\hat{H}_\mathrm{S}$, the unitary part of~\eqref{eq:master} remains trivially unchanged. The superoperator $\hat{\mathfrak{L}}_t$ nevertheless carries an explicit $t$-dependence through the coefficients $\kappa(t)$, $\Gamma(t)$, $h(t)$ and $f(t)$. A shift $t\to t+\tau$ therefore takes $\hat{\mathfrak{L}}_t \to \hat{\mathfrak{L}}_{t+\tau}\neq\hat{\mathfrak{L}}_t$, and the master equation is \textbf{not time-translation-invariant}. This breaking has a trivial origin: the factorized initial condition selects a preferred instant, and the transient relaxation of the bath toward its stationary state induces memory effects that depend on the elapsed time. In the late-time Markovian limit, where the coefficients saturate to constants, ordinary time-translation invariance returns, with the dressed Hamiltonian as its generator. The boost violation derived below remains structurally distinct, persisting even in this Markovian limit.

\emph{Galilean boosts}\textemdash The boost generator for the system reads $\hat{\boldsymbol{G}}_\mathrm{S} = M_\mathrm{S}\rs - \ps t$. Under $\hat{U}_u = e^{\mathrm{i}\boldsymbol{u}\cdot\hat{\boldsymbol{G}}_\mathrm{S}}$, direct evaluation gives $\rs \to \rs - \boldsymbol{u}\,t$ and $\ps \to \ps - M_\mathrm{S}\boldsymbol{u}$. The Hamiltonian line picks up a non-trivial shift, $\hat{H}_\mathrm{S}'-\hat{H}_\mathrm{S} = -\boldsymbol{u}\cdot\ps + \tfrac{1}{2}M_\mathrm{S}\boldsymbol{u}^{2}$; the anomalous-diffusion and normal-diffusion lines transform invariantly on their own, since their $c$-number shifts vanish inside the nested commutators with $\rs$. The time-derivative inhomogeneity $\mathrm{d}\hat{U}_u/\mathrm{d}t\,\hat{U}_u^{\dagger}=-\mathrm{i}\boldsymbol{u}\cdot\ps$ in Eq.~\eqref{eq:master_transf} exactly cancels the Hamiltonian-line shift (SM~\cite{SM}).

The sole surviving contribution comes from the \emph{dissipative} line. Under the boost, $\acomm{\ps - M_\mathrm{S}\boldsymbol{u}}{\hat{\rho}_\mathrm{S}'} = \acomm{\ps}{\hat{\rho}_\mathrm{S}'} - 2M_\mathrm{S}\boldsymbol{u}\,\hat{\rho}_\mathrm{S}'$. The cross term yields
\begin{equation}\label{eq:boost_Break}
\frac{\mathrm{d}\hat{\rho}_\mathrm{S}'}{\mathrm{d}t} = \hat{\mathfrak{L}}_t\,\hat{\rho}_\mathrm{S}'(t) + 2\mathrm{i}M_\mathrm{S}\Gamma(t)f(t)\,\boldsymbol{u}\cdot\comm{\rs}{\hat{\rho}_\mathrm{S}'}.
\end{equation}
The master equation is \textbf{not boost-invariant}. The symmetry-breaking term scales directly with the damping coefficient $\Gamma(t)f(t)$, which also governs the exponential decay of the system's mean momentum. This establishes an exact correspondence: \emph{boost invariance breaks if and only if the environment damps the system's momentum}. For any realistic spectral density at finite temperature, $\Gamma(t)f(t) \neq 0$, and boost invariance fails.

\emph{Spectral origin and the fluctuation--dissipation theorem}\textemdash The cancellation pattern admits a transparent algebraic--spectral interpretation. Of the four operator structures in Eq.~\eqref{eq:master}, the first three either transform invariantly on their own (the anomalous-diffusion and normal-diffusion lines, whose $c$-number shifts vanish inside the nested commutators with $\rs$) or generate a Hamiltonian-line shift that is exactly canceled by the $\mathrm{d}\hat{U}_g/\mathrm{d}t$ terms of Eq.~\eqref{eq:master_transf}. The commutator-anticommutator structure of the dissipative line, by contrast, couples the $c$-number shift in $\ps$ to the outer commutator with $\rs$ and survives those mechanisms as the inhomogeneous residual of Eq.~\eqref{eq:boost_Break}. Operator structure---not spectral origin---selects which terms remain covariant.

At the spectral level, $\delta\Omega^{2}(t)$ derives from the dispersive (real) part of the bath response, $\Gamma(t)$ and $\Gamma(t)f(t)$ are sourced from the friction kernel $\eta(s)$---the absorptive part of the susceptibility ($\chi''$)---and $\Gamma(t)h(t)$ from the noise kernel $\nu(s)$~\cite{Hu1992,Grabert1988}; at finite times each coefficient is a convolution of the kernels with the classical homogeneous solutions, with the assignment exact in the Markovian limit. The dissipative coefficient is therefore tied to the absorptive bath response, paralleling the QFT distinction between virtual dressing (renormalization, symmetry-preserving) and real momentum absorption (dissipation, symmetry-breaking).

At thermal equilibrium, the fluctuation--dissipation theorem (FDT) ties the noise and friction kernels in a definite ratio set by the thermal factor $\coth(\beta\hbar\omega/2)$~\cite{Grabert1988,Pachon2019}: the same absorptive bath spectrum that feeds $\Gamma(t)f(t)$ also feeds, via FDT, the diffusive coefficient $\Gamma(t)h(t)$. Setting $\Gamma(t)f(t)=0$ to restore boost covariance therefore requires either eliminating the bath's momentum-exchange channel altogether or violating the FDT in equilibrium. The implication chain runs in one direction: bilinear-coupled Galilean invariance, combined with the kinetic structure $\hat{H}_\mathrm{S}\supset\ps^{2}/2M_\mathrm{S}$, forces $[\hat{H}_\mathrm{S},\hat{V}_\text{int}]\neq 0$ and hence $\Gamma(t)f(t)\neq 0$ whenever the bath spectral density has support on the relevant frequencies; FDT then forbids the \emph{independent} suppression of $\Gamma f$ at fixed thermal noise $\Gamma h$, since both descend from the same absorptive bath spectrum.

\emph{Scope of the covariant-map framework}\textemdash The broader literature recognizes that tracing out degrees of freedom can break symmetries of the full system, e.g., in the context of reference frames and environment-induced superselection~\cite{Bartlett2007, Zanardi2001}. The present results delineate the regime in which the program initiated by Holevo~\cite{Holevo1996} and extended by Gasbarri \emph{et al.}~\cite{PhysRevLett.119.100403} admits a microscopic foundation in a bilinearly coupled environment. Covariance of the dynamical map, $\mathcal{M}_t = \hat{U}_g^\dagger \circ \mathcal{M}_t \circ \hat{U}_g$, requires the system Hilbert space to represent the symmetry in its entirety. Because the full Galilean boost generator is extensive (mass-weighted), the system-only generator $\hat{\boldsymbol{G}}_\mathrm{S}$ carries only a fraction $M_\mathrm{S}/M$ of the total; the partial trace irretrievably discards the environmental share. The microscopic derivation presented here establishes that, for any physical system subject to dissipation, this kinematic obstruction is a strict no-go: no Galilean-invariant Caldeira--Leggett-type microscopic model yields a boost-covariant reduced map whenever $\Gamma(t)f(t)\neq 0$. The result thereby identifies, within the bilinear-coupling class, the precise regime in which non-Markovian covariant extensions~\cite{Smirne2010} and macroscopicity measures admit a microscopic foundation, and the regime in which their predictions deviate from the underlying Hamiltonian dynamics. The restriction to bilinear-coordinate couplings is essential: master equations arising from non-bilinear microscopic mechanisms (e.g., scattering-mediated bath couplings) can yield boost-covariant reduced dynamics by construction and lie outside the present scope.

\emph{Quantitative estimates and regime of validity}\textemdash Equation~\eqref{eq:boost_Break} provides a precise measure of symmetry breaking. A natural dimensionless figure of merit emerges from the ratio of the symmetry-breaking energy scale to the thermal energy $k_\mathrm{B} T$. For an Ohmic bath at high temperatures, $\Gamma(t) \to \gamma$ and $f(t)$ scales to leading order as $\hbar \gamma / k_\mathrm{B} T$~\cite{Hu1992, Pachon2019}, with subleading Matsubara corrections of order $(\hbar\gamma/k_\mathrm{B} T)^{2}$ analyzed in the SM~\cite{SM}. The relevant dimensionless parameter controlling the structural violation is therefore $\hbar \gamma / k_\mathrm{B} T$.

\begin{figure}[t]
\centering
\includegraphics[width=\columnwidth]{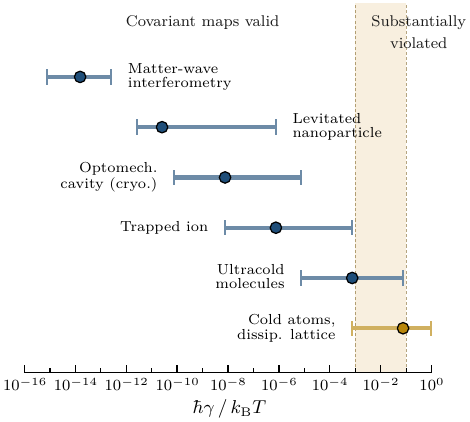}
\caption{Dimensionless boost-violation parameter $\hbar\gamma/k_\mathrm{B} T$ across representative open-quantum-system platforms in the high-temperature Ohmic regime. Markers indicate representative values quoted in the text; horizontal bars span the experimental ranges of $\gamma$ and $T$~\cite{Arndt2014, Gonzalez-Ballestero2021, Aspelmeyer2014, Leibfried2003, Bohn2017, Grynberg2001}. The shaded band delineates the crossover between the decoherence-dominated regime (left, covariant maps serve as an excellent approximation) and the dissipation-dominated regime (right, where Eq.~\eqref{eq:boost_Break} attains quantitative significance). Cold atoms in dissipative optical lattices span the crossover and enter the violated regime; ultracold molecules reach its lower edge. Bars cap at $\hbar\gamma/k_\mathrm{B} T = 1$, beyond which the leading-order high-temperature expansion for $f(t)$ ceases to apply (SM~\cite{SM} for the expansion-error analysis).}
\label{fig:violation}
\end{figure}

Figure~\ref{fig:violation} surveys representative open-quantum-system platforms on this scale. A typical macroscopic optomechanical levitated nanoparticle at room temperature ($T = 300$~K, $\gamma \sim 10^3$~Hz) yields $\hbar \gamma / k_\mathrm{B} T \sim 10^{-10}$: the violation remains entirely negligible and Galilean-covariant maps serve as an excellent approximation. Cold atoms in dissipative optical lattices~\cite{Grynberg2001} ($T \sim 1~\mu$K, $\gamma \sim 10^4$~Hz) instead reach $\hbar\gamma/k_\mathrm{B} T \sim 10^{-1}$. There the covariance assumption omits the dissipative sector by construction; while this poses no problem for short-time decoherence bounds, applications that integrate the dynamics over times comparable to $\gamma^{-1}$---including long-time extrapolations of macroscopicity measures~\cite{Nimmrichter2013}---should explicitly assess the missing dissipative contribution. Covariant maps therefore qualify as a controlled approximation only when the experimental time scale stays extremely short compared to the momentum relaxation time ($t \ll \gamma^{-1}$), where decoherence dominates over dissipation.

Two operational consequences for experiment follow. First, the regime $\hbar\gamma/k_\mathrm{B} T \sim \mathcal{O}(1)$ is precisely the regime in which residual deviations from a covariant-map ansatz fitted to short-time data calibrate $\Gamma f$ in a manner independent of any covariant template, complementary to standard thermometry through the FDT. Cold atoms in dissipative optical lattices and ultracold molecules~\cite{Grynberg2001} thus offer a natural setting for an operational test of the no-go derived here. Second, the converse statement constrains fundamental physics: a system observed to retain exact boost covariance at the reduced level despite finite thermal contact would imply either a bath structure with no momentum-exchange channel or a violation of the FDT.

\emph{Generality beyond the Caldeira--Leggett model}\textemdash The Caldeira--Leggett model serves as the starting point because it provides the most general analytical treatment of quadratic system--bath couplings, and the cancellation pattern just established applies to the full class of quadratic-map applications that follow naturally from this model. Crucially, the stochastic decomposition of the influence functional~\cite{Stockburger2002, Koch2008, Diosi1998}---which generalizes the diffusion term to a stochastic equation for the reduced state---extends the boost-violation mechanism beyond the quadratic regime: this generalization constitutes a central conceptual contribution of the present work. For a system with arbitrary potential $V(\hat{x})$ bilinearly coupled to a bath, the reduced density matrix takes the form of a stochastic average over complex noise realizations, $\hat{\rho}_\mathrm{S}(t) = \mathcal{M}\!\left[|\Psi_1(t)\rangle\langle\Psi_2(t)|\right]$. Under a boost, a stochastic force $\xi(t)$ coupling linearly to $\hat{x}$ transforms as $\xi(t)\hat{x} \to \xi(t)(\hat{x} + ut)$, and the $c$-number piece $\xi(t)\,ut$ shifts the phase of each stochastic sample. The imaginary part of the friction kernel $L''(t)$---which encodes the bath's causal response, i.e.\ dissipation---introduces an asymmetric phase weighting between $|\Psi_1\rangle$ and $\langle\Psi_2|$, manifest as the opposite signs of the $\pm\nu(t)$ noise terms in the underlying stochastic Schr\"odinger equations~\cite{Koch2008}. The asymmetry enters through the second-cumulant (variance/correlation) structure of the noise rather than its mean, so the standard linear-cumulant cancellation that ordinarily averages such phases away does not apply, and a residual, non-covariant deformation of $\hat{\rho}_\mathrm{S}$ survives the average $\mathcal{M}$. The diffusive (stochastic) sector therefore carries the boost-violation mechanism for arbitrary $V(\hat{x})$~\cite{Grabert1988}. A rigorous algebraic proof for arbitrary anharmonic potentials remains an open problem; the present work frames this extension as a conjecture supported by physical argument.

\emph{Pure dephasing and Galilean geometry}\textemdash A pure dephasing interaction commuting with the system Hamiltonian, $[\hat{H}_\mathrm{S},\hat{V}_{\text{int}}]=0$, automatically yields $\Gamma f=0$. Galilean invariance, however, requires $\hat{V}_{\text{int}}$ to depend only on relative coordinates, $\hat{V}_{\text{int}}\propto(\rs-\rn)^2$; combined with the kinetic term $\ps^{2}/2M_\mathrm{S}$ in $\hat{H}_\mathrm{S}$, this forces $[\hat{H}_\mathrm{S},\hat{V}_{\text{int}}]\neq 0$. Pure dephasing is therefore incompatible with Galilean-invariant bilinear coupling: a Galilean-invariant relative-coordinate coupling forbids the bath cloud from remaining purely virtual, since the kinetic term ensures that dispersive ``dressing'' of system parameters is always accompanied by an absorptive momentum-exchange channel.

\emph{Initial correlations}\textemdash Initial system--bath correlations can fundamentally alter the operator structure of the exact master equation~\cite{Stelmachovic2001, Talkner2006}. The imaginary part of the bath force--force correlation $L''(t)$ that determines $\Gamma(t)f(t)$ is fixed by the spectral density alone and remains insensitive to the choice of initial state~\cite{Hu1992, Grabert1988}; as long as the bath can exchange momentum with the system, the dissipative channel stays open. Combined with the kinematic fact that $\hat{\boldsymbol{G}}_\mathrm{S}$ generates only the system's share $M_\mathrm{S}/M$ of the total boost, any term encoding irreversible momentum transfer must transform inhomogeneously, so boost-breaking should persist in the presence of initial correlations---though a fully rigorous operator-level proof for arbitrary correlated initial states remains open.

\emph{Driving and the effective restoration of boost covariance}\textemdash A natural follow-up question asks whether the boost-breaking term in Eq.~\eqref{eq:boost_Break} can be dynamically suppressed without modifying the bath. Parametric driving of the system---through time-modulation of $\omega_\mathrm{S}$ or of an internal coupling---introduces a squeezing-rate scale $\mu$. In two parametrically coupled harmonic oscillators in contact with independent thermal baths, this rate competes with the bath-induced decoherence rate $\gamma\,k_\mathrm{B} T/\hbar\omega$: the standard quantum limit $\hbar\omega/k_\mathrm{B} T>1$ for nonequilibrium-entanglement survival renormalizes to $\mu/\gamma > k_\mathrm{B} T/\hbar\omega$~\cite{Galve2010}, further relaxed by non-Markovian dynamics through an effective scaling of $T$, $\gamma$, and $\mu$~\cite{Estrada2015}. The boost-breaking coefficient at high temperature scales as $\Gamma f\sim\hbar\gamma^{2}/k_\mathrm{B} T$, so suppressing the boost-breaking phase $\Gamma f\,t$ over a driving cycle $t\sim\mu^{-1}$ requires $\mu \gtrsim \hbar\gamma^{2}/k_\mathrm{B} T$. The Galve--Pach\'on--Zueco threshold dominates this boost-suppression threshold by a factor $(k_\mathrm{B} T)^{2}/(\hbar^{2}\omega\gamma) \gg 1$ in the high-$T$, weak-coupling regime, so satisfying the entanglement-survival condition automatically suppresses the boost-breaking dynamics over a driving cycle. This implication runs in one direction: nonequilibrium steady-state entanglement is \emph{sufficient} for effective boost-covariance restoration on the relevant timescale, not necessary.

\emph{Conclusions}\textemdash Starting from a system--environment model whose Lagrangian is manifestly Galilean invariant, the analysis above demonstrates that the exact master equation for the reduced dynamics breaks boost invariance. The violation stems entirely from the dissipative (anticommutator) term and vanishes if and only if the damping coefficient $\Gamma(t)f(t)$ is zero. Within the harmonic Caldeira--Leggett class with factorized initial conditions, this constitutes a strict no-go: \emph{Galilean boost invariance and dissipation are mutually exclusive}.

The Caldeira--Leggett model serves as the natural starting point because it provides the most general analytical platform for quadratic system--bath couplings, and the operator-level result obtained here transfers to the full class of quadratic-map applications. The stochastic decomposition of the influence functional then extends the boost-violation mechanism beyond the quadratic regime, with the diffusive (stochastic) sector itself carrying the obstruction for arbitrary anharmonic potentials. The extension to correlated initial states, supported by physical arguments rather than a fully algebraic proof, is framed as a conjecture above. Equivalently, for any non-trivial bath spectral density, the master equation forces a choice between bilinear-coupled Galilean invariance, the FDT at thermal equilibrium, and reduced boost covariance---only two of which can simultaneously hold, with consequences that connect the symmetry analysis presented here to quantum thermodynamics.

Parametric driving offers a one-directional escape: the high-temperature entanglement scenarios of Refs.~\cite{Galve2010,Estrada2015} sit in a parameter regime where the squeezing rate is large enough to also suppress the boost-breaking dynamics over a driving cycle, suggesting that nonequilibrium steady-state entanglement and effective boost covariance share a common dynamical origin. The same extensive (mass-weighted) structure of the Poincar\'e boost generator suggests that the kinematic obstruction identified here plausibly carries over to the relativistic setting, with the analogous microscopic analysis left for future work.

\emph{Acknowledgments}\textemdash The R+D+I efforts from guane Enterprises supported this work.

\end{document}